\documentstyle[12pt,psfig]{article}
        \def\be{\begin{equation}}
        \def\ee{\end{equation}}

\begin{document}
\begin{titlepage}
\vspace*{5mm}

\begin{center} {\large \bf Partially Asymmetric Simple Exclusion Model  
in the Presence of an Impurity on a Ring} \\

\vskip 1cm

\centerline {\bf Farhad H Jafarpour \footnote {e-mail:JAFAR@theory.ipm.ac.ir}} 
\vskip 1cm
{\it  Department of Physics, Sharif University of Technology, }\\
{\it P.O.Box 11365-9161, Tehran, Iran }\\
{\it  Institute for Studies in Theoretical Physics and Mathematics,}\\
{\it P.O.Box 19395-5531, Tehran, Iran}
\end{center}

\begin{abstract}

We study a generalized two-species model on a ring. The original model [1]
describes ordinary particles hopping exclusively in one direction in the
presence of an impurity. The impurity hops with a rate different from that of 
ordinary particles and can be overtaken by them. Here we let 
the ordinary particles hop also backward with rate $q$. 
Using the Matrix Product Ansatz (MPA), we obtain the relevant quadratic 
algebra. A finite dimensional representation of this algebra enables us to 
compute the stationary bulk density of the ordinary particles, as well as
the speed of impurity on a set of special surfaces of the parameter space. 
We will obtain the phase structure of this model in the accessible region 
and show how the phase structure of the original model is modified.
In the infinite-volume limit this model presents a shock in one of 
its phases.

\end{abstract}



\end{titlepage}

\newpage

\section{Introduction}

Recently much attention has been focused on one-dimensional reaction-diffusion
processes. These models can describe many physical phenomena 
such as hopping conductivity, growth processes and traffic flows [2-4]. 
They are also of interest from the mathematical point of view due to their   
relation to integrable quantum chain Hamiltonians [5,6].
The simplest model of this kind is the Asymmetric Simple Exclusion Process 
(ASEP) with open boundaries [7]. This model comprise particles which jump 
independently to their right with hard core repulsion along a one-dimensional
lattice. The open boundary conditions mean that particles are injected at one
end of the lattice and are removed at the opposite end.
This model exhibits a shock structure in the density profile of particles. 
In periodic boundary conditions case the microscopic location of the shock can be 
identified by defining a second class particle (impurity). 
The ASEP in the presence of an impurity on a ring has been studied in 
the two following cases: \\

1) In the first case, the single impurity hops in the opposite direction 
relative to the ordinary particles [8,9]. In this case a first-order phase transition 
between a low-density and a traffic jam phase is observed . \\

2) In the second case, the impurity moves in the same direction 
as the ordinary particles [1,17]. The phase diagram of this model consists of six
distinctive phases (I-VI) in which two of them are symmetric to other phases
under a charge conjugation and reflection symmetry. The authors have shown 
that one phase exists in the system in which the impurity causes a shock. \\

Another example of such driven diffusive systems is the Partially Asymmetric
Simple Exclusion Process (PASEP). In this model the particles are allowded to jump
both to their immediate right (with rate $p$) or left (with rate $q$) site, if the target site is not already occupied. This model has 
been extensively studied both with open boundaries and on a ring [10]. 

In this paper we will study the effects of the presence of a single impurity on 
the PASEP on a ring. Here the ordinary particles can hop to their immediate right
(left) site, provided that it is empty, with rate $ 1 $ $(q \leq 1)$. The single 
impurity can only hop to its immediate right site (if it is not already occupied) 
with rate $\alpha$$(\leq 1)$ and can be exchanged from the left with the ordinary 
particles with rate $\beta$$(\leq 1)$. For $q=0$ this model reduces to the model
(2) as discussed above. 
Using Matrix Product Ansatz (MPA) introduced in [11], we obtain the relevant
quadratic algebra which has both finite and infinite dimensional 
representations. For simplicity, we adopt a finite dimensional representation  
of the algebra and carry out all the calculations using a grand
canonical ensemble in which the population of ordinary particles 
can fluctuate. By adjusting the fugacities of the ordinary particles 
and of the holes, one can produce some fixed densities for them. 
Although the finite dimensional representation restricts us to the 
region under the surface $ \alpha + \beta + q = 1 $ in the three-dimensional 
parameter space, nevertheles we shall find the exact phase 
structure and calculate precisely the density profile 
of ordinary particles and the speed of impurity in this region.
With this exact results, we will show that three phases exist in this
region. In two of them, which are symmetric 
to each other under a charge conjugation and reflection symmetry, 
the density profile of particles has an exponential behavior.
We will determine the relevant correlation lengths in these phases and the critical
values of the rates that characterize the divergence of these correlation 
lengths. In the third phase the density profile of the ordinary particles 
is linear which is a signature of a shock.                                         

This paper is organized as follows. In section 2 we will describe the model. 
In section 3 we will write the weights of the configurations in the stationary state 
in terms of a trace of $L+1$ non-commuting matrices and obtain the quadratic
algebra. In section 4 we will introduce all possible representations of the 
quadratic algebra and discuss different phases in the accessible region
of the phase space. In section 5 using one and two-dimensional 
representations of the algebra we will obtain some exact results for an infinite
system. In the last section we will compare our results with those obtained in
[1] for $q=0$.

\section{The Model}

The model proposed here contains two species of particles on 
a closed ring of $L+1$ sites which are labled from $1$ to $L+1$. 
We specify each configuration of the system by an $L+1$-tuple 
$( \tau_1 , \tau_2, ..., \tau_{L+1} )$, where $\tau_i=1$ if site $i$ is
occupied with a particle of kind $1$, $\tau_i=2$ if it is occupied with 
a particle of kind $2$, and $\tau_i=0$ if site $i$ is empty. There are $M$ 
particles of kind $1$, and only one particle of kind $2$ on the ring. We will
refer to them as ordinary particles and impurity respectively.\\
The system evolves under stochastic dynamics. The possible exchanges between
two adjacent sites during a time interval $dt$ is as follows:\\

\begin{eqnarray}
10 \rightarrow 01 \ \ \  with \ \ \  rate \ \ \  1  \nonumber \\
01 \rightarrow 10 \ \ \  with \ \ \  rate \ \ \  q   \nonumber \\
20 \rightarrow 02 \ \ \  with \ \ \  rate \ \ \  \alpha \nonumber \\
12 \rightarrow 21 \ \ \  with \ \ \  rate \ \ \  \beta  \nonumber  
\end{eqnarray}

The space of configuration is connected. Each configuration can evolve into any other and
therefore, it has a unique stationary state [18]. 
As we mentioned above, we will perform all the calculations in the grand canonical ensemble
and let the number of ordinary particles fluctuate around a mean value.
We will show that in the large $L$ limit the fluctuations 
in the density of ordinary particles drop to zero so that in this limit,
the results of the canonical ensemble agree with those of the grand canonical ensemble.\\
The above process can be mapped onto the one introduced in [12] by 
interchanging the impurity and vacancies ( 2 $\Longleftrightarrow $ 0). 
It can also be considered as a special case of the three-species diffusion 
problems introduced in [13].

\section{The Stationary Measure and the Quadratic Algebra}  

According to the matrix product formalism, the 
stationary probability $P(\{ \tau \})$ of any configuration $ \{ \tau \}$
can be written as a trace of a product of non-commuting operators. Since this
model is translationally invariant, one can always keep the single impurity  
at site $L+1$ and write the normalized weight $P( \{ \tau \} )$ 
in terms of three operators $D$, $E$ and $A$:\\

\be
P(\{ \tau \})=P(\{ \tau_1, \tau_2,...,\tau_L,\tau_{L+1}=2 \})=
{1 \over Z_L} Tr[ \prod_{i=1}^{L} (x \tau_i D+y(1-\tau_i)E)A]
\ee

The non-zero real variables, $x$ and $y$, are the fugacities of
ordinary particles and vacancies respectively. In the grand canonical ensemble approach,
one must choose the value of the fugacities to fix the density of ordinary particles 
to be $ \rho = { M \over L }$. Although it would be sufficient to introduce only
one fugacity (
since the total number of sites  " $L+1$"  is fixed), we use both of them to make the
symmetry between ordinary particles and vacancies more apparent.
The normalization factor $Z_L$ in the denominator of Eq.$(1)$,
which plays a role analogus to the partition function in equilibrium statistical
mechanics, is a fundumental quantity and can be calculated using the fact 
$ \sum_{ \{ \tau \} } P(\{ \tau_1,...,\tau_L,\tau_{L+1} \})=1$.
Thus one finds \\

\be
Z_L = \sum_{all \ \ configurations} Tr[ \prod_{i=1}^{L} (x \tau_i D+y(1-\tau_i)E)A] =
Tr(C^{L}A)
\ee \\
in which $C=xD+yE$. The operators $D$, $E$ and $A$ satisfy the following quadratic algebra:

\begin{eqnarray}
DE-qED & = & D+E \\
{\beta}DA & = & A \\
{\alpha}AE & = & A 
\end{eqnarray}

Assuming that $ A= \vert V \rangle \langle W \vert $,
Eqs.(3-5) can be written in terms of
$\vert V \rangle $ and $ \langle W \vert $ as follows

\begin{eqnarray}
DE-qED & = & D+E \\
D \vert V \rangle & = & { 1 \over \beta} \vert V \rangle \\
\langle W \vert E & = & { 1 \over \alpha} \langle W \vert 
\end{eqnarray}

With this assumption, the calculation of $Tr(...)$ in expression (2) reduces to
the calculation of a matrix element

\be 
Z_L=\langle W \vert C^L \vert V \rangle 
\ee

In the next section we will investigate all possible representations 
of (6-8).

\section{Representations of the Quadratic Algebra}

In [14] the Fock representation of the most general quadratic algebra has been 
obtained. It has been shown that in order to have a finite dimensional representation
of certain quadratic algebras, the parameters of the model, say $\alpha$ ,$\beta$ and
$q$ in (6-8), should satisfy a set of constraints.  \\
The algebra (6-8) has a one-dimensional representation for $ \alpha, \beta, q < 1$, 
where the operators $D$, $E$ and $A$ are represented by real numbers. Here the
vectors $ \vert V \rangle $ and $\langle W \vert $ can be discarded and one is
dealing with a scalar product state in (9). This representation exists if

\be
D={ 1 \over \beta} \ \ , \ \ E= { 1 \over \alpha}  \ \ , \ \ A=1
\ee \\
and the following constraint holds

\be
\alpha + \beta + q = 1
\ee
In Fig.1 we have ploted (11) in the three-dimensional parameters space. 
One can easily check that the following matrices [15]

\be {D} ={1 \over 1-q} 
\left( \begin{array}{lllll} 
1+a&0&0&.&.\\
0&1+aq&0&.&.\\
0&0&1+aq^2&0&.\\
.&.&.&.&.\\
.&.&0&1+aq^{n-2}&0\\
.&.&0&0&1+aq^{n-1}\\ 
\end{array} \right )
\ee \\

\be {E} ={1 \over 1-q} 
\left( \begin{array}{lllll} 
1+{1 \over a} &0&0&.&.\\
1&1+{1 \over aq}&0&.&.\\
0&1&1+{1 \over aq^2}&0&.\\
.&.&.&.&.\\
.&.&1&1+{1 \over aq^{n-2}}&0\\
.&.&0&1&1+{1 \over aq^{n-1}}\\ 
\end{array} \right )
\ee \\
and the vectors

\be 
\vert V \rangle = \left( \begin{array}{c}
1 \\ 0 \\ . \\ . \\ . \\ 0 
\end{array} \right)  \ \ \ , \ \ \ 
\vert W \rangle = \left( \begin{array}{c}
1 \\ \omega_2 \\ \omega_3 \\ . \\ . \\ \omega_n 
\end{array} \right) \ \ \  
\omega_i=\prod_{j=0}^{i-2} {1 \over a} 
 ( { 1\over q^{n-1}} -{ 1 \over q^j}) \ \ , \ \ i=2,...,n 
\ee \\
in which 

\be 
a= {1 - q - \beta \over \beta} = 
{ 1 \over { { 1-q- \alpha \over \alpha} } q^{n-1} }
\ee \\
constitute an $n$-dimensional representation of (6-8) for $ \alpha, \beta , q < 1$, 
provided that the following constraint between the parameters $ \alpha$, $\beta$
and $q$ holds

\be
({1-q- \beta \over \beta}) ({1-q- \alpha \over \alpha})=q^{1-n} 
\ee

Note that for $n=1$ the above constraint reduces to (11). Using (16) it can be 
verified that for $q < 1$, the region of the phase space which is accessible 
by the totaliy of all finite dimensional $(n \geq 2)$ representations is

\be
\alpha + \beta + q<1
\ee

In Fig.2 we have plotted (16) for two values of $n$. One can see that as
the dimension of representation increases from $2$, the two-dimensional 
surface (16) begins to approach towards the $q$ axis. Therefore the finite dimensional
representation
of the algebra (6-8) allows us to derive exact results for this model only on
some special surfaces given by (16). However using the same conjecture
proposed in [15], one can introduce a kind of analytical continuation to obtain
results valid in the whole accessible region given by (17).

In what follows we show that in the region specified by (17), only three phases
can exist and this result is independent of $n$ (dimension of the representation).
We now find the asymptotic behavior of (9) for large values of $L$. 
Note that the fugacity of ordinary particles $x$ and vacancies $y$ 
remain constant in this limit. In the thermodynamic limit
$( L \longrightarrow \infty)$, the behavior of $Z_L$ is governed by the
largest eigenvalue of $C$. The eigenvalues $ \xi_i$ of

\be {C} = {xD+yE} = $$
$${1 \over 1-q} 
\left( \begin{array}{cccc} 
(x+y)+xa+{y \over a} &0&.&.\\
y&(x+y)+xaq+{y \over aq}&0&.\\
.&.&.&.\\
.&0&y&(x+y)+xaq^{n-1}+{y \over aq^{n-1}}\\ 
\end{array} \right )
\ee \\
can readily be computed

\be
\xi_i={ 1 \over 1-q} \{ (x+y)+xaq^{i-1}+{ y \over aq^{i-1}} \} \ \ , \ \ i=1,...,n
\ee

We notice that all the eigenvalues of $C$ lie on the curve [15]

$$ z \longrightarrow { 1 \over 1-q }( (x+y)+xz+ {y \over z}) $$\\
in which $z=a,aq,...,aq^{n-1}$. Therefore apart from the 
values of the fugacities, the largest eigenvalue $\xi_{max}$
of $C$ takes one of the following values: \\

$I)\ \ \xi_{max} = \xi_1$, if $\xi_1 > \xi_n$ or  equivalently 
  $ x( { 1-q- \beta \over \beta} ) > y( { 1-q- \alpha \over \alpha} )$ \\

$II)\ \ \xi_{max} = \xi_n$, if $\xi_n > \xi_1$ or  equivalently 
  $ x( { 1-q- \beta \over \beta} ) < y( { 1-q- \alpha \over \alpha} )$     \\

$III)\ \ \xi_{max} =\xi_1=\xi_n$, if $ \xi_1=\xi_n $ or equivalently 
    $ x( { 1-q- \beta \over \beta} ) = y( { 1-q- \alpha \over \alpha} )$    \\

These three different cases correspond to different phases.
The process is invariant when the direction 
of motion is reversed and the following transformations are applied

\begin{eqnarray}
Particle \ \ of \ \ kind \ \ 1 \ \ & \longrightarrow & \ \  
Particle \ \ of \ \ kind \ \ 0 \ \ (Vacancy) \nonumber \\
Site \ \ number  \ \ i \ \ & \longrightarrow & \ \ Site \ \ number \ \ L+1-i \nonumber \\
\rho \ \ & \longrightarrow & \ \ 1-\rho \nonumber  \\
\alpha \ \ &  \longrightarrow & \ \ \beta  \nonumber
\end{eqnarray}

Now considering that the interchange of $x$ and $y$ is equivalent to 
the exchange of the density of
ordinary particles and vacancies, $(I)$ is symmetric to $(II)$ under 
these transformations.\\

By using (18) one can see that in the regions $(I)$ and $(II)$,
$C$ is diagonalizable, so we expect that 
all correlation functions of form $ \langle \tau_{i_1}...\tau_{i_{L+1}} \rangle $ 
depend exponentially on the distances involved. In the region $(III)$
the eigenvalues of $C$ coincide

\be
\xi_k=\xi_{n-k+1}=
{1 \over 1-q} \{(x+y)+y({1-q-\alpha \over \alpha})(q^{k-1}+q^{n-k}) \}
\ \ , \ \ k=1,...,n
\ee \\
In this case, since $C$ has an off-diagonal line (see 18),
it is not diagonalizable. Hence it implies an algebraic behavior of all correlation functions
(see discussions in [16]). In the next section we will show this 
explicitly for a two-dimensional representation.\\
The algebra (6-8) also has infinite dimensional representations [13] and by using them
one has access to the entire phase space without any constraints on the parameters.
Since for these representations the calculations seem to be very difficult, 
we shall adopt one and two-dimensional representations to study the
general behavior of some interesting quantities, such as the density profile
of ordinary particles and the speed of impurity in different phases.\\

\section{One and Two Dimensional Representations. Exact Results}

As mentioned in the previous section, using $n$-dimensional representations
$(n \geq 2)$, only the region $\alpha + \beta + q <1 $ is accessible,
where three phases exist. One-dimensional representation (10)
limits us to that part of phase space given by the constraint (11).
On this two dimensional surface, the partition function (9) has a simple form \\

\be
Z_L=({ x \over \beta} + { y \over \alpha } )^{L}
\ee \\
The mean value of density of the ordinary particles which is defined
by\\

\be
\rho = { x \over L } { d \over dx} \ln {Z_L} 
\ee  \\
can easily be evaluated

\be
\rho = { { x \over \beta} \over { {x \over \beta} + { y \over \alpha} } }
\ee \\
If $\rho_i$ denotes the expectation that site $i$ is occupied with an
ordinary particle in the stationary state, knowing that the only impurity
is at site $L+1$, we have \\

\be
\rho_i = x{ {Tr(C^{i-1}DC^{L-i}A)}  \over  {Tr(C^{L}A)}  }=x
{
{ \langle W \vert C^{i-1} D C^{L-i} \vert V \rangle } \over
{ \langle W \vert C^L \vert V \rangle }
}
\ee \\
Consequently, using (10) and (23), we are able to evaluate the density profile of ordinary
particles in this case

\be
\rho_i = \rho
\ee  \\
In Fig.3 we have plotted the phase diagram of the model for a constant $q$.
The accessible region (17) lies under the line $\alpha + \beta =1-q$. On this
line each configuration has the same stationary probability and mean field 
results become exact. 
In the stationary state, the speed of impurity $v$ can be expressed as

\be
v=y \alpha { Tr( E C^{L-1} A) \over Tr(C^L) } -x \beta
 { Tr(C^{L-1} D A) \over Tr(C^L) } \\
 =(y-x){ Z_{L-1} \over Z_L }
\ee \\
when use has been made of the algebraic relations (6-8). 
Using (21), (23) and (26) the speed of impurity is found to be

\be
v=\alpha-(1-q) \rho
\ee

One may also note that for $ \alpha > (1-q) \rho $, (27) becomes
negative i.e. the impurity starts moving backward. This can easily be understood
from the fact that whenever an ordinary particle appears behind the impurity,
it overtakes the impurity while pushing it backward (See section 2). Thus, when
the density of ordinary particles exceeds a critical value
$ \rho_c={ \alpha \over 1-q}$, a macroscopic negative current of impurity
evolves in the system.\\

Now we will consider a two-dimensional representation. The following matrices

\be {D} =
\left( \begin{array}{ll} 
{1 \over \beta}&{ i \over \sqrt{q}}\\
0&{ \beta+q \over \beta} \\
\end{array} \right ) 
\ \  , \ \ {E} =
\left( \begin{array}{ll} 
{1 \over \alpha}&0 \\
{ i \over \sqrt{q}}&{ \alpha+q \over \alpha} \\
\end{array} \right )
\ee \\
with the vectors

\be \vert V \rangle= \left( \begin{array}{c}
1 \\ 0  
\end{array} \right) \ \ ,\ \
\vert W \rangle= \left( \begin{array}{c} 1 \\ 0  \end{array} \right)
\ee   \\
satisfy (6-8), if the following relation holds

\be
(q + \alpha)(q + \beta)=q
\ee  

For $n=2$, this representation is equivalent to (12-14) and (30) concides
with (16). As we saw, the properties of the matrix \\

\be   
{C} =
\left( \begin{array}{cc} 
{ { x \over \beta } + { y \over \alpha} } & { xi \over \sqrt{q} } \\
{ yi \over \sqrt{q} } &
{ x( { \beta+q \over \beta } ) + y ( { \alpha+q \over \alpha } )} \\
\end{array} \right)  
\ee \\
is of prime importance in determining the phase structre of the model.
The eigenvalues of $C$ can easily be computed \\

$$
\xi_1={ x \over \beta } + ( 1 + { q \over \alpha } ) y \  \ , \  \
\xi_2=(1 + { q \over \beta} ) x + { y \over \alpha }
$$ \\

These eigenvalues also coincide with those obtained from (19) for $n=2$.
For $ \xi_1 \neq \xi_2 $, one can introduce

\be {U} =
{ 1 \over \sqrt{
 { i \over \sqrt q } ( y ( { 1-q-\alpha \over \alpha } )
                      - x ( { 1-q-\beta \over \beta} ) ) } }
\left( \begin{array}{cc}
{xi \over \sqrt{q}} & {-y{1-q-\alpha \over \alpha} } \\
{i \over \sqrt{q}} & { -{1-q-\beta \over \beta} } \\
\end{array} \right) 
\ee \\
to diagonalize $C$

$$
{U C U^{-1}} =
\left( \begin{array}{ll} 
{\xi_1}&{0}\\
0&{ \xi_2} \\
\end{array} \right ) $$
\\
and using (9) and (29) obtain the following expresion for $Z_L$ 

\be
Z_L= {
{ x ( {1-q-\beta  \over  \beta} ) \xi_{1}^{L}  - 
   y ( {1-q-\alpha \over \alpha} ) \xi_{2}^{L} }  \over
   { x ( { 1-q-\beta \over \beta }) - y ( {1-q-\alpha \over \alpha} ) } }
\ee

In the thermodynamic limit, two cases can be distinguished.
The first case, corresponding to the phase I, is specified with $ \xi_1 > \xi_2$,
where the first term in (33) becomes dominant.
The mean density of ordinary particles (22) can be evaluated as

\be
\rho = {  
{x  \over \beta }  \over
{ {x \over \beta} + { y (1 + { q \over \alpha })}   }  }
\ee  

The inequality $ \xi_1 > \xi_2 $ can be written in terms of $\rho$ as

$$ \alpha >  (1-q){(1 - \rho)} $$ \\
which with $ \alpha + \beta + q < 1 $ define the boundaries of the phase $(I)$ (See
Fig.3).
In the $ q \rightarrow 0 $ limit, this phase corresponds to the phase (VI)
of the model studied in [1]. Using (24), (28-30), (32) and (34) we obtain

\be
\rho_1= 1- { 1 \over q+ \alpha}(1- \rho) \ \ , \ \
\rho_L= \rho
\ee     

The density profile decreases exponentially from $ \rho_1$ to $ \rho_L$.
One can define a characteristic length measuring the range of the effect
of the impurity

\be 
\xi^{-1} = \ln { q+\alpha \over 1+\rho(q-1) } 
\ee \\

Note that the correlation length diverges as 
$ \rho \longrightarrow 1-{\alpha \over 1-q} $. For $ L >> \xi$ the density 
profile of the ordinary particles is of the form

\be
\rho_i =
c_1e^{-i \over \xi} + c_2e^{ i-L \over \xi}+ c_3
\ee     \\
in which

\begin{eqnarray}
c_1 & = & { {-xy ({1-q-\alpha \over \alpha}) ({1-q-\alpha \over \alpha}
-{1-q-\beta \over \beta}) }
\over { ( y {1-q-\alpha \over \alpha} - x {1-q-\beta \over \beta} ) \xi_{2}}} \nonumber \\
c_2 & = & { { x(y-x) } \over
{ q { (y{1-q-\alpha \over \alpha}-x{1-q-\beta \over \beta}) \xi_{1} } }} \nonumber \\
c_3 & = & { x { x({1 \over q} - {1 \over \beta}{ 1-q-\beta \over \beta })+y{(1-q-\alpha)}
({ \beta+q \over \beta } ) \over
{{q{(y{1-q-\alpha \over \alpha}-x{1-q-\beta \over \beta} )\xi_{1}} }}}} \nonumber 
\end{eqnarray}

Using (26), (33) and (34) we find the following expression for the speed
of impurity in phase $(I)$

\be
v= { \alpha \over q+ \alpha} -(1-q) \rho
\ee

One observes that for $ \rho > { 1-q-\beta \over 1-q} $, the average speed
of the single impurity again becomes negative. It can easily be checked
that in $ q \rightarrow 0 $ limit the values of $ \rho_1 $, $ \rho_L$
and $v$ approach to their corresponding values in the phase (VI) of [1].
Next we will examine the second phase
which is characterized with $ \xi_2 > \xi_1 $. This phase is symmetric to
phase $(I)$ and all the results can be obtained using the
transformations introduced in section 4.
Finally the third phase $(III)$ occurs when $\xi_1=\xi_2$, where
$C$ is not diagonalizable (See 32).
Nevertheless, $Z_L$ can be computed noting that in this case $C$ can be written as

\be {C} =
\left( \begin{array}{cc} 
{{x \over \beta}+{y \over \alpha} }&
{ xi \over \sqrt{q}}\\
{ iy \over \sqrt{q}}&
{ x({ \beta+q \over \beta }) + y({ \alpha+q \over \alpha}) } \\
\end{array} \right )= $$
$$({ x \over \beta }+{ y \over \alpha }
-x({ 1-q-\beta \over \beta }))I+
\left( \begin{array}{cc} 
{ x({1-q-\beta \over \beta}) } & { xi \over \sqrt{q}} \\
{iy \over \sqrt{q}} & {-y({1-q- \alpha \over \alpha})}\\
\end{array} \right ) =
({ x \over \beta }+{ y \over \alpha }
-x({ 1-q-\beta \over \beta }))I+S 
\ee \\
in which

\be
S^2=0
\ee \\
Using this property with (9) and (29) yield

\be
Z_L=\langle W \vert C^L \vert V \rangle=
( x ( {q+\beta \over \beta} )  + { y \over \alpha } )^{L+1}
\{ x( { q+\beta \over \beta} ) + { y \over \alpha } +
L x [ { 1-q-\beta \over \beta} ] \}
\ee \\
and the mean value of density of ordinary particles (22) is found to be

\be
\rho=
{ { \sqrt{q} x + { 1 \over 2 } (1+q) \sqrt{xy} } \over
{ \sqrt{q} x +\sqrt{q} y + (1+q) \sqrt{xy} }}
\ee \\
The boundaries of the phase $(III)$ (See Fig.3) are limited to
\begin{eqnarray}
\alpha &<&  (1- \rho)(1-q) \nonumber \\
\beta  &<&  \rho (1-q)  \nonumber
\end{eqnarray} \\
which can be distinguished for $ q < 1 $ using the fact that
$x( { 1-q- \beta \over \beta} ) = y( { 1-q- \alpha \over \alpha} )$,
(30) and (42). The density profile of ordinary particles, as mentioned above,
has an algebraic behavior in this phase and increases linearly from 
$ \rho_1= { \beta \over {1-q}} $ to $ \rho_L= 1-{ \alpha \over {1-q} } $ 
according to

\be
\rho(z)={ \beta \over 1-q } + ( 1- { \beta + \alpha  \over 1-q } )z \ \ , \ \
0 \leq z \leq 1
\ee 

For $q=0$ this phase corresponds to phase $(V)$ in [1] where a shock structure
has been observed for certain values of the density $\rho$. The linear profile
which is the consequence of a fluctuating shock front, has also been 
observed in ASEP and PASEP with open boundaries [11,14,15]. Here the relation 
$x( { 1-q- \beta \over \beta} ) = y( { 1-q- \alpha \over \alpha} )$ prevents
fixing the fugacities $x$ and $y$ and we cannot adjust the density $\rho$
given by (42) to the desired value. Therefore we cannot see a real shock
structure in the grand canonical ensemble. Nevertheless one can observe it in 
this phase using the canonical ensemble [19]. 
The speed of impurity (26) in this phase is of the form

\be
v= \alpha -\beta
\ee \\
which is independent of $ \rho$. Note that for $\beta > \alpha $ the speed of
impurity again becomes negative. One can check that in this phase the
values of $ \rho_1$, $\rho_L$ and $v$ approach their corresponding
values in the phase (V) in $ q \rightarrow 0 $ limit. \\
The fluctuation in the density of ordinary particles [17] which is given by

\be
[ { x \over L} { d \over dx} ( { x \over L} {d \over dx} \ln{Z_L} ) ]^{1 \over 2}
\ee \\
can also be calculated in each phase. Using (33) and (41) one can easily
check that in large $L$ limit these fluctuations drop to zero like
$ 1 \over \sqrt{L}$ therefore the results of the grand 
canonical ensemble agree with those obtained from the canonical one.

\section{Comparison and Concluding Remarks}

As we mentioned in section 2, this model can be considered
as a simple generalization of the model studied in [1]. Here we compare the
results obtained there (we call it the model A) with thoes in our model (the
model B).\\
For $q=0$, the quadratic algebra of model B reduces to the one of model A. 
Using an infinite dimensional representation of this algebra, the
author has solved model A exactly and shown that it possesses six
distinctive phases (I-VI). Since we have used a finite dimensional representation
of the quadratic algebra (6-8), only the region $ \alpha + \beta + q < 1 $ lies
in the accessible region where three phases exist. These phases (I-III) 
correspond to the phases (VI), (IV) and (V) in model A respectively.
In model A, the density profile of the ordinary particles has an exponential behaviour 
in the phases (VI) and (VI) but it has a shock structure in phase (V).  
In phases (I) and (II) of model B, the density profile has also an exponential 
behaviour but with a modified correlation length. In the region (III) we have obtained 
a linear profile for the density of the ordinary particles in the grand canonical ensemble
which indicates the presence of a shock in this phase. 
From the phase structure point of view, the backward hopping of the
ordinary particles does not change the number of phases at least in the 
accessible region and only the co-existance lines are shifted.\\ 
In this paper we have used a finite dimensional representation of the
algebra and done all the calculations in grand canonical ensemble which 
make the calculations rather simple. It would be interesting to study 
the region $ \alpha + \beta + q > 1 $, especially the shock structure in the
phase (III), using an infinit-dimensional representation of (6-8)
in the canonical ensemble. \\


{ \large \bf Acknowledgement: } \\

I would like to express my gratitute to K. Mallick for his valuable 
hints and comments through an E-mail correspondence.
I'm also grateful to M. Barma for useful discussion and
V. Karimipour for critical reading of the manuscript.


\newpage
\begin{center}
{\large Figure Captions}
\end{center}
\begin{description}
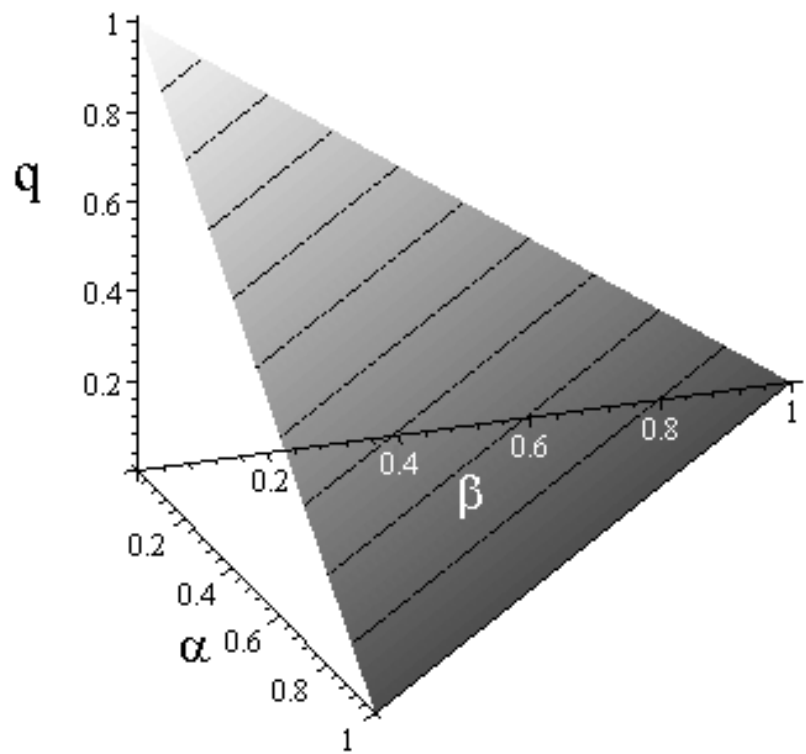
\item[Fig.\ 1] Plot of the equation (11).
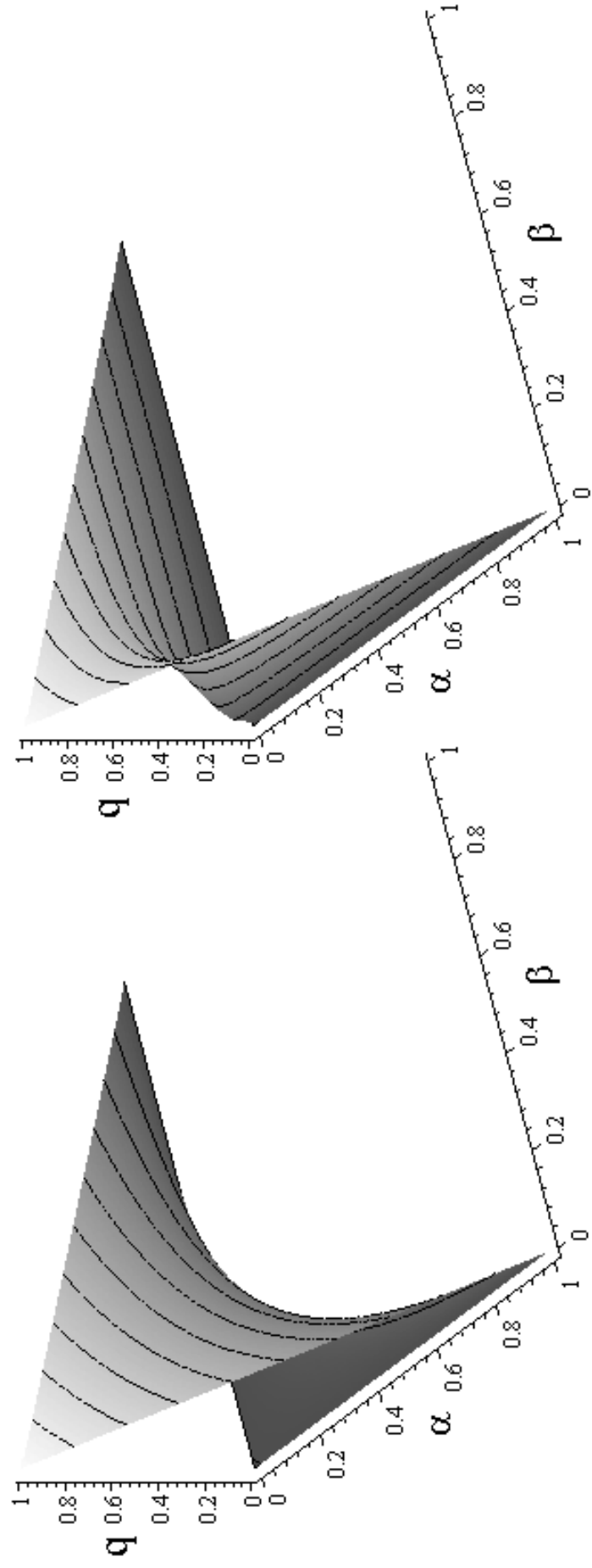
\item[Fig.\ 2] Plot of the equation (16) for n=2 (Left) and n=5 (Right).
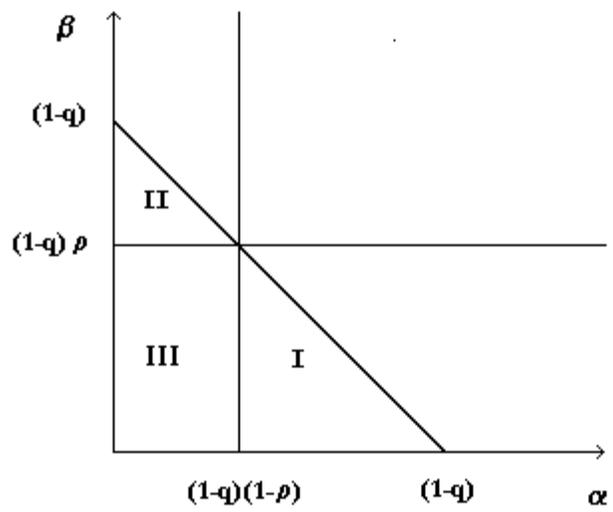
\item[Fig.\ 3]The phase diagram.
\end{description}

\end{document}